\documentclass[aps,prl,twocolumn,groupedaddress,superscriptaddress,preprintnumbers]{revtex4-1}

\usepackage{graphicx}
\begin{document}


\title{Photo-thermal response in dual-gated bilayer graphene}


\author{M.-H. Kim}
\affiliation{Center for Nanophysics and Advanced Materials,
Department of Physics, University of Maryland, College Park,
Maryland 20742, USA}
\author{J. Yan}
\affiliation{Center for Nanophysics and Advanced Materials,
Department of Physics, University of Maryland, College Park,
Maryland 20742, USA} \affiliation{Physics Department, University
of Massachusetts, Amherst, MA 01002, USA}
\author{R. J. Suess}
\affiliation{Institute for Research in Electronics and Applied
Physics, University of Maryland, College Park, Maryland 20742,
USA}
\author{T. E. Murphy}
\affiliation{Institute for Research in Electronics and Applied
Physics, University of Maryland, College Park, Maryland 20742,
USA}
\author{M. S. Fuhrer}
\affiliation{Center for Nanophysics and Advanced Materials,
Department of Physics, University of Maryland, College Park,
Maryland 20742, USA}
\author{H. D. Drew}
\affiliation{Center for Nanophysics and Advanced Materials,
Department of Physics, University of Maryland, College Park,
Maryland 20742, USA} \email{hdrew@umd.edu}


\begin{abstract}
The photovoltaic and bolometric photoresponse in gapped bilayer
graphene was investigated by optical and transport measurements. A
pulse coincidence technique at 1.5 $\mu$m was used to measure the
response times as a function of temperature. The bolometric and
photovoltaic response times were found to be identical implying
that the photovoltaic response is also governed by hot electron
thermal relaxation. Response times of $\tau \sim$ 100 - 20 ps were
found for temperatures from 3 - 100 K. Above 10 K, the relaxation
time was observed to be $\tau$ = 25 $\pm$ 5 ps, independent of
temperature within noise.
\end{abstract}

\pacs{78.67.Wj, 78.47.D-, 78.56.-a}

\maketitle


There is growing recognition that graphene has exceptional
potential as a new optoelectronic material, which has led to a
flurry of recent research activity and rapid
advances.~\cite{Avouris2009, Avouris2010, Novoselov2011}
Graphene's unique massless band structure gives rise to direct
transitions and strong (specific) coupling to light at all
wavelengths,~\cite{CastroNeto2009} and ultra-fast response (from
nanosecond to femtosecond)~\cite{CastroNeto2009} room temperature
operation for many applications. A photovoltaic response has been
observed for visible light and we have recently observed both
photovoltaic and bolometric response in bilayer graphene at THz
frequencies.~\cite{Jun_Kim2012} Diode-like rectification behavior
is observed with contacts to dissimilar
metals.~\cite{GoldhaberGordon2008, Avouris2009,
Avouris2010,Jun_Kim2012} However, the mechanism of the
photovoltaic response has not been definitively identified. Both
p-n junction physics similar to conventional semiconductor
photovoltaic sensors and a thermoelectric mechanism remain viable
possibilities.  In a recent study we observed a hot electron
bolometric response in bilayer graphene, which highlighted the
outstanding thermal properties of graphene.~\cite{Jun_Kim2012}
Therefore, understanding the role of hot electron effects in the
photoresponse of graphene is critical to the development of
graphene-based optoelectronic devices such as bolometers and
photovoltaic sensors.~\cite{Gabor2011}

Excited electrons in graphene thermalize quickly on the
femtosecond time scales~\cite{Rana2008, Heinz2010} by
electron-electron scattering.~\cite{DasSarma2008} These hot
electrons transfer their thermal energy to the graphene lattice by
the emission of phonons on a much longer time scale because of the
weak electron-phonon interaction.~\cite{DasSarma2008, Fuhrer2008,
MacDonald2009, Viljas2010} The thermal relaxation of hot electrons
by optical phonons~\cite{Rana2011, Heinz2010, SunXu2012} and by
acoustic phonons~\cite{Jun_Kim2012, McEuen2012} has received much
recent attention. The processes of cooling by optical and acoustic
phonons are clearly distinguishable because their thermal
timescales differ by a few orders of magnitude. High pulse energy
radiation produces hot electrons with energies above the optical
phonon energy ($\sim$ 200 meV) that cool by optical phonon
emission in a timescale of a few picoseconds.~\cite{Rana2011,
Heinz2010, SunXu2012} For longer times and/or lower pulse energy
radiation, acoustic phonon assisted cooling is dominant with
sub-nanosecond timescales.~\cite{Jun_Kim2012, McEuen2012}

Hot electrons can be utilized for bolometric and photovoltaic
photoresponse detection.~\cite{Jun_Kim2012, Levitov2011,
Gabor2011} The bolometric response makes use of the temperature
dependence of the resistivity, which is significant in gapped
bilayer graphene. On the other hand, the hot electrons can also
give rise to a photo-thermoelectric response. Diffusion of heat
and carriers to the contacts produce a thermoelectric response. A
competing mechanism for photovoltaic response is charge separation
by the built-in electric fields at metal-graphene junctions due to
proximity doping. It remains unclear which of these two mechanisms
dominates in graphene photovoltaic devices.

In this paper, we use electrical transport and optical
photoresponse measurements to characterize the bolometric and
photovoltaic response of a dual-gated bilayer graphene device. The
temperature-dependent resistance of the device which allows a
bolometric response is characterized both optically and with AC
transport measurements, which together establish that the response
is thermal. We found that light also generates a voltage across
the sample with zero bias current. We compare this photovoltaic
response with the well-understood bolometric signal in the same
device as functions of dual-gate voltages and temperature. In
particular pulse coincidence measurement reveals that the photo
voltage displays the same temperature-dependent relaxation time as
the bolometric response, demonstrating that diffusive hot carrier
relaxation in graphene dominates the observed photo voltage of the
device.

\begin{figure}
\includegraphics[scale=0.4, bb = 10 0 527 603, clip =
true]{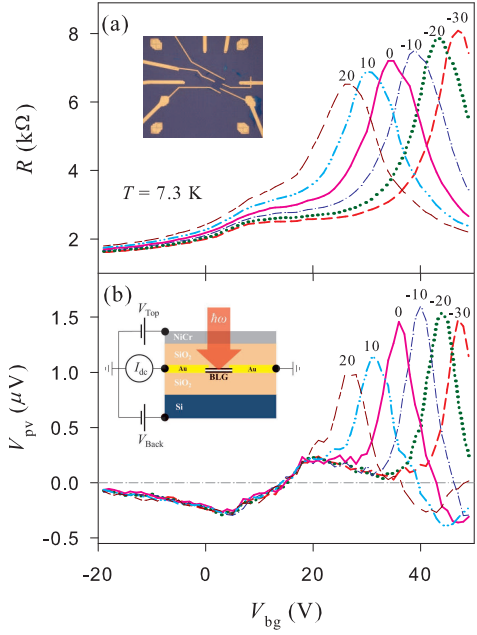} \caption{\label{Fig1}\textbf{Photovoltaic response
and resistance of a dual-gated bilayer graphene.} (a) Resistance
and (b) photovoltaic response as a function of back gate voltage
for different top gate voltage at $T = 7.3$ K and zero bias
current. Inset in (a) shows an optical micrograph of the bilayer
graphene device. Inset in (b) shows schematic of device geometry
and electric-field-effect gating.}
\end{figure}

The bilayer graphene device we studied was fabricated by
mechanical exfoliation of natural graphite on a resistive silicon
wafer (200 $\Omega$cm) which was ion implanted with boron and
annealed to provide a highly conducting but transparent back-gate
at cryogenic temperatures. A 300 nm thick SiO$_2$ layer was grown
by a dry oxidation process from the silicon wafer. A micrograph of
the device is shown in Fig.~\ref{Fig1}(a) inset and schematic of
device geometry is shown in Fig.~\ref{Fig1}(b) inset. A thin
Nichrome film was used as a semitransparent top-gate. Details of
the device structure and the gating scheme can be found in
Ref.~\onlinecite{Jun_Kim2012}. The dual-gated structure allows for
independent tuning of carrier density and bandgap of the bilayer
graphene device. Figure \ref{Fig1}(a) shows the device resistance
$R$ at 7.3 K as a function of back gate voltage $V_{bg}$ at
various top gate voltages $V_{tg}$. A broad resistance peak
appears near $V_{bg}$ = 10 V independent of $V_{tg}$ and is
attributed to the part of the bilayer device that is not gated by
the top gate.~\cite{Jun2010} The other sharper peak shifts with
$V_{tg}$ and is attributed to dual-gated device region.

The photoresponse and sample resistance were measured
simultaneously. The photoresponse shown in Fig.~\ref{Fig1}(b) was
measured with bias current $I_{dc}$ = 0 which gives the
photovoltaic response $V_{pv}$. At $V_{bg}$ = 10 V where the broad
peak of $R$ occurs, $V_{pv}$ also doesn't depend on $V_{tg}$, and
$V_{pv}$ crosses zero at $V_{bg} \approx$ 15 V. This behavior is
similar to that observed in photo-thermoelectric results reported
in graphene.~\cite{McEuen2010, Gabor2011, SunXu2012} For $V_{bg}
>$ 20V, $V_{pv}$ depends on both $V_{tg}$ and $V_{bg}$, and reaches a maximum
value at the maximum $R$.

To measure the response times of these signals we used a pulse
coincidence technique. The photoresponse was studied at 1.56
$\mu$m with a pulsed laser with a 65-fs pulse width and 100 MHz
repetition rate. Pulses from two fiber lasers are locked together
with a tunable time separation at repetition rate near 100 MHz
(Menlo Systems), which allows pulse coincidence measurements with
precise time delays from a few ps to 10 ns without a mechanical
delay line. The absorption of 1.56 $\mu$m radiation in the
graphene was estimated to be 1.2\% by considering effects due to
the silicon substrate and the Nichrome top
gate.~\cite{Jun_Kim2012} The graphene absorbs an average power of
0.37 nW from the pump and probe pulses and generates a temperature
rise $\Delta T$, which can be measured using the temperature
dependence of $R$.

\begin{figure}
\includegraphics[scale=0.4, bb = 10 0 527 673, clip =
true]{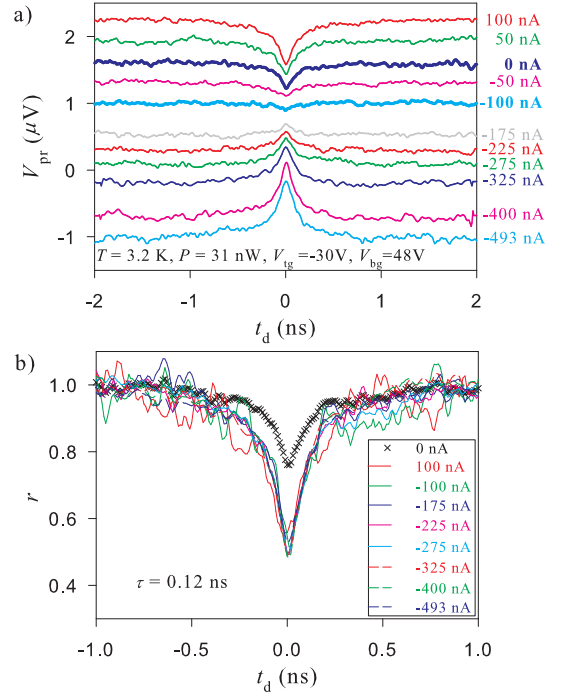} \caption{\label{Fig2}\textbf{Bias current
dependence of the pump-probe measurements.} Photoresponse from
pump-probe laser pulses as a function of time delay $t_d$ at 3.2 K
and laser absorbed power of 31 nW. The sample is gated to charge
neutrality with $V_{tg}$ = -30 V, $V_{bg}$ = 48 V. (a) Bias
current $I_{dc}$ dependence of probe-induced photoresponse voltage
$V_{pr}(t_d)$. The $I_{dc}$ = 0 curve is the photovoltaic
response. (b) $r$ is normalized bolometric response $\Delta
V(I_{dc}) = V(I_{dc}) - V(0)$ and photovoltaic response $V(0)$. $r
= V_{pr}(t_d)/V_{pv}^0$, where $V_{pv}^0 = V_{pv} (t_d \gg \tau)$.
The thermal response time $\tau$ is defined as the half-width at
half-maximum of the dip. All dips have a similar time constant
$\tau = 0.12 \pm 0.01$ ns.}
\end{figure}

The dependence of the photoresponse with pulse time delay for
different bias currents is shown in Fig.~\ref{Fig2}(a) under
conditions that the device is gated to its charge neutral point.
At zero bias current the signal is purely photovoltaic. For
non-zero bias there is also a bolometric signal given by $V_b =
I_{dc} \Delta R$ which was reported earlier.~\cite{Jun_Kim2012} We
find that the total photoresponse can be described as
$V_{pr}(I_{dc}) = V_b(I_{dc}) + V_{pv}$ allowing a separation of
the photovoltaic and bolometric contributions. Therefore the
bolometric response can be obtained from $V_b(I_{dc}) =
V_{pr}(I_{dc}) - V_{pr}(0)$. It is seen in the figure that this
bolometric response is dominant except near $I_{dc} = 0$ where the
response is purely photovoltaic.

These pulse time delay data allow a measurement of the response
time $\tau$ of the two components of the photoresponse. For long
pulse delay times, $t_d$, average probe-pulse induced photo
voltage, $V_{pr}$ does not change with respect to the time delay
$t_d$. When the delay is short ($t_d < \tau$), however, the
magnitude of $V_{pr}$ is reduced due to the nonlinear radiation
power dependence of the response so that the photo voltage
$V_{pr}(t_d)$ displays a peak or dip at $t_d = 0$. The magnitude
of this peak or dip increases with the non-linear power dependence
of $V_{pr}$.

Figure \ref{Fig2}(b) shows $V_{pv}$ and $V_b(I_{dc})$ normalized
to the response in the absence of the pump pulse for several
different $I_{dc}$. All of the normalized $V_b$ for different
$I_{dc}$ collapse to one curve because the small Joule heating
$I_{dc}^2R$ does not significantly raise the electron temperature
above the lattice temperature. Both bolometric and photovoltaic
responses produce a narrow dip with a sub-nanosecond width at zero
time delay when the pump and probe pulses are coincident.
Surprisingly, the widths of both bolometric and photovoltaic dips
are seen to be the same to within the experimental error. The time
constants determined by the half widths at half maximum of the
dips are $0.12 \pm 0.01$ ns. This demonstrates that both $V_{pv}$
and $V_b(I_{dc})$ have the same response time and since the
bolometric response is clearly thermal~\cite{Jun_Kim2012} this
implies that the photovoltaic response is also thermal. Similar
results are observed at $V_{tg} = 0$ V and $V_{bg} = 30$ V for
temperatures below 10 K.

\begin{figure}
\includegraphics[scale=0.35, bb = 10 0 587 833, clip =
true]{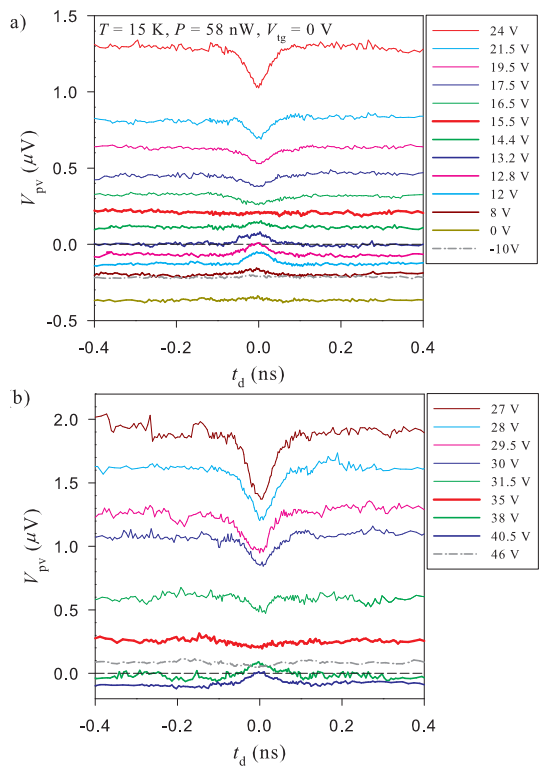} \caption{\label{Fig3}\textbf{Gate voltage
dependence of the pump-probe measurements.} Pump-probe pulse
induced photovoltaic response as a function of time delay at 15 K
and laser power of 58 nW. $I_{dc} = 0$. The data was taken at
several back gate voltages $V_{bg}$ with zero top gate voltage (a)
below and (b) above $V_{bg} = 25$ V where the maximum photovoltaic
response is found. The thin dashed line at $V_{bg} = 0$ is a guide
line. All cusps have the same thermal time constant $\tau = 25 \pm
5$ ps within error.}
\end{figure}

To gain further insight into the nature of the photo voltage, we
measured its gate voltage dependence. Figure \ref{Fig3} shows back
gate voltage dependence of photo voltage at $T = 15$ K with
$V_{tg} =$ 0 and $I_{dc} =$ 0. As can be seen from the data in
Fig.~\ref{Fig1}(a), the top gate does not gate the entire device.
To obtain uniform gating we control only the back gate voltage
with zero top gate voltage. Figures \ref{Fig3}(a) and (b) display
the photovoltaic response below and above the maximum $V_{pv}$
observed at around $V_{bg} = 25$ V. The peak or dip structure is
associated with the sign of $V_{pv}$, and its depth or height
depends on the nonlinear power dependence of $V_{pv}$. Both sign
and power nonlinearity depend on back gate voltage. For example,
at $V_{bg} = 15.5$ V the response $V_{pv}(t_d)$ is independent of
$t_d$ indicating that $V_{pv}$ is linear with radiation power. As
the power nonlinearity of $V_{pv}$ grows above or below $V_{bg} =
15.5$ V, the dip or peak of $V_{pv}$ appears and grows.
Remarkably, however, all of the pump-probe data have the same
$\tau = 25$ ps $\pm 5$ ps. The gate-independent time constant
shows that the photovoltaic response is thermal at all gate
voltages not only at maximum $R$ with respect to $V_{bg}$ as shown
in Fig.~\ref{Fig2}(b) where it could be directly compared with the
bolometric response. This observation demonstrates that the
photovoltaic response time in bilayer graphene is the intrinsic
thermal time constant of hot electron energy relaxation.

\begin{figure}
\includegraphics[scale=0.45, bb = 10 0 527 410, clip =
true]{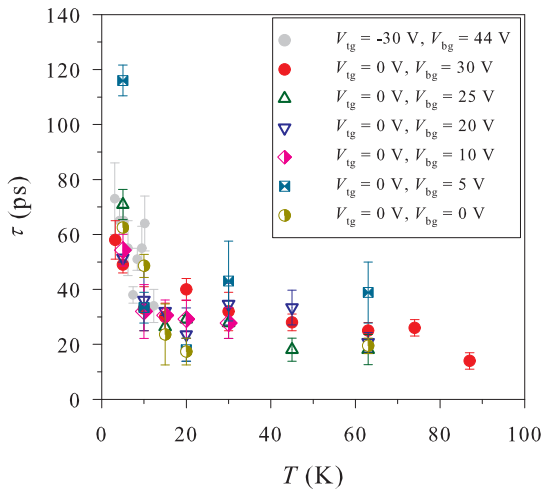} \caption{\label{Fig4}\textbf{Temperature
dependence of thermal response time.} Thermal time constant $\tau$
of the photovoltaic response vs. temperature measured by the pulse
coincidence technique for several different dual-gate voltage
settings.}
\end{figure}

We also measured the temperature dependence of the response time
in our graphene device. Figure \ref{Fig4} exhibits $\tau$ obtained
from the photovoltaic response in the temperature range 3 K - 87 K
at several different dual-gate voltages. The response time for
different gate voltages coincides within error as shown in
Fig.~\ref{Fig4}. The time constant decreases from $\sim$ 80 ps at
3 K to $\sim$ 20 ps at 80 K. Above $T \sim$ 10 K, $\tau$ is seen
to be temperature independent to within experimental error.

The thermal relaxation rate is given by the ratio of the
electronic heat capacity $C$ to thermal conductance $G$. The
thermal conductance was obtained using transport measurements as
described in Ref.~\onlinecite{Jun_Kim2012}. For $T < 8$ K, the
transport measurements gives $G = 0.5 \times (T/5)^{3.45}$ nW/K
which is in reasonable agreement with the value estimated for
cooling by acoustic phonons.~\cite{Viljas2010} A crossover of the
thermal conductance from $T^3$ to linear $T$ is predicted for $T >
T_{BG}$, where $T_{BG}$ is the Bloch-Gr\"{u}nheisen temperature
given by $k_B T_{BG} \approx 2h v_s k_F$.~\cite{Viljas2010}
Assuming a sound velocity $v_s = 2.6 \times 10^4$
m/s~\cite{Efetov2010} and a disorder-induced charge density of
$n_{\rm rms} \sim 10^{12}$ cm$^{-2}$~\cite{Yan2008}, we find
$T_{BG} \sim 70$ K. Although our sample is nominally charge
neutral at $R_{\rm max}$, it is widely accepted that disorder
creates electron-hole puddles~\cite{Martin2008} and thus $T_{BG}$
is non-zero at all gate voltages. Transport measurements show that
the Bloch-Gr\"{u}nheisen regime behavior occurs for $T < 0.2
T_{BG} \sim 14$ K.~\cite{Efetov2010} The behavior of $G$ and $C$
above $T \sim 0.2 T_{BG}$ may be complicated by disorder induced
supercollision cooling~\cite{McEuen2012,Levitov2012} and/or the
non-parabolic band structure of gapped bilayer
graphene~\cite{CastroNeto2009} which leads to small Fermi
energies. We measured $G = 0.91 \times T^{1.04}$ nW/K for $T
> 8$ K, which is reasonable in view of these considerations.

On the other hand, diffusion cooling of hot electrons also gives a
linear $T$ dependent thermal conductance. Diffusion cooling
provides a thermal conductance $k = \Lambda T/R_g$, by the
Wiedermann-Franz law, where $\Lambda = 24.4$ nW$\Omega$/K$^2$ is
the Lorentz number, and $R_g$ is bilayer graphene resistance. At
the peak resistance for our sample, $k = 3.4 \times 10^{-12}
\times T^{1.0}$ W/K, which is two orders of magnitude smaller than
the electron-phonon conductance. We conclude that acoustic phonon
mediated cooling of hot electrons is dominant in our samples.

We note, however, that the thermoelectric and photovoltaic signals
are a consequence of diffusion. For asymmetric contacts the
thermal diffusion and charge flow at the two contacts differs
leading to a net potential difference. The diffusion length $\xi =
(k/G)^{1/2}$ is estimated to be 0.5 $\mu$m at 10 K which is much
smaller than the sample size of 5 $\mu$m so that the sample
temperature rise and response time is dominated by the thermal
conductance to the lattice which greatly reduces the
thermoelectric signals in these large area, low conductance
samples. The thermoelectric effect produces an electric field $e =
S \nabla T$, where $S$ is the Seebeck coefficient and $\nabla T$
is the gradient of the temperature. The heat conductance to the
lattice diminishes $\nabla T$ at the contact and therefore the
thermoelectric field by a factor of $2\xi/L$ compared with pure
diffusion.

At low temperatures ($k_BT < \mu$, where $\mu$ is the local Fermi
energy in the graphene and $k_B$ is the Boltzmann constant) the
electronic specific heat is $C = \alpha T$, where $\alpha =
(\pi^2/3)v(E_F)k_B^2$, where $v(E_F)$ is the density of states for
bilayer graphene.  In the parabolic band approximation of
(ungapped) graphene $v(E_F) \approx \gamma_1/(\pi \hbar^2 v_F^2)$
where the interlayer coupling $\gamma_1 = 390$
meV~\cite{Zhang2008}, $v_F = 1 \times 10^6$ m/s is the monolayer
graphene Fermi velocity. For our sample area of 25 $\mu m^2$, this
gives $\alpha = 2.6 \times 10^{-20}$ J/K$^2$. Thus the thermal
response time of our bilayer sample can be estimated $\tau = C/G
\approx 29$ ps independent of temperature for $T > 8$ K which is
in reasonable agreement with the measured $\tau$ shown in
Fig.~\ref{Fig4}.

In summary, we have reported photovoltaic response time
measurements on gapped bilayer graphene. The devices show both
bolometric and photovoltaic responses, which were separated by
their bias current dependence. The identical response time
constants observed for the bolometric and photovoltaic responses
as a function of gate voltages and temperature implies that both
effects are governed by the same intrinsic hot electron-phonon
thermal relaxation process. The observed response times of 10 -
100 ps indicates that hot electron relaxation occurs through
acoustic phonon emission. These observations support the growing
realization that graphene appears to have great promise for fast
sensitive photo detectors over a wide spectral range.

\begin{acknowledgments}
The authors thank A. B. Sushkov, G. S. Jenkins, and D. C. Schmadel
for helping with the optical cryostat setup and for valuable
discussions. This work was supported by IARPA, the ONR MURI
program and the NSF (grants DMR-0804976 and DMR-1105224) and in
part by the NSF MRSEC (grant DMR-0520471).
\end{acknowledgments}



\begin{thebibliography}{24}

\bibitem{Avouris2009} F. Xia, T. Mueller, Y.-M. Lin, A. Valdes-Garcia, and
P. Avouris, Nature Nanotech. \textbf{4}, 839 (2009).

\bibitem{Avouris2010} T. Mueller, F. Xia, and P. Avouris, Nature Photonics \textbf{4},
297 (2010).

\bibitem{Novoselov2011} T. J. Echtermeyer, L. Britnell, P. K. Jasnos, A. Lombardo,
R. V. Gorbachev, A. N. Grigorenko, A. K. Geim, A. C. Ferrari, and
K. S. Novoselov, Nature Commun. \textbf{2}, 458 (2011).

\bibitem{CastroNeto2009} A. H. Castro Neto, F. Guinea, N. M. R. Peres, K. S.
Novoselov, and A. K. Geim, Rev. Mod. Phys. \textbf{81}, 109
(2009).

\bibitem{Jun_Kim2012} J. Yan, M.-H. Kim, J. A. Elle, A. B. Sushkov, G. S.
Jenkins, H. M. Milchberg, M. S. Fuhrer, and H. D. Drew, Nature
Nanotech. \textbf{7}, 472 (2012).

\bibitem{GoldhaberGordon2008} B. Huard, N. Stander, J. A. Sulpizio, and D. Goldhaber-
Gordon, Phys. Rev. B \textbf{78}, 121402 (2008).

\bibitem{Gabor2011} N. M. Gabor, J. C. W. Song, Q. Ma, N. L. Nair, T. Taychatanapat,
K. Watanabe, T. Taniquchi, L. S. Levitov, 5 and P.
Jarillo-Herrero, Science \textbf{334}, 648 (2011).

\bibitem{Rana2008} P. A. George, J. Strait, J. Dawlaty, S. Shivaraman,
M. Chandrashekhar, F. Rana, and M. G. Spencer, Nano Lett.
\textbf{8}, 4248 (2008).

\bibitem{Heinz2010} C. H. Lui, K. F. Mak, J. Shan, and T. F. Heinz, Phys.
Rev. Lett. \textbf{105}, 127404 (2010).

\bibitem{DasSarma2008} W.-K. Tse, E. H. Hwang, and S. D. Sarma, Appl. Phys.
Lett. \textbf{93}, 023128 (2008).

\bibitem{Fuhrer2008} J.-H. Chen, C. Jang, S. Xiao, M. Ishigami, and M. S.
Fuhrer, Nature Nanotech. \textbf{3}, 206 (2008).

\bibitem{MacDonald2009} R. Bistritzer and A. H.MacDonald, Phys. Rev. Lett. \textbf{102},
206410 (2009).

\bibitem{Viljas2010} J. K. Viljas and T. T. Heikkil{\"{a}}, Phys. Rev. B \textbf{81}, 245404
(2010).

\bibitem{Rana2011} J. H. Strait, H. Wang, S. Shivaraman, V. Shields,
M. Spencer, and F. Rana, Nano Lett. \textbf{11}, 4902 (2011).

\bibitem{SunXu2012} D. Sun, G. Aivazian, A. M. Jones, J. S. Ross, W. Yao,
D. Cobden, and X. Xu, Nature Nanotech. \textbf{7}, 114 (2012).

\bibitem{McEuen2012} M. W. Graham, S.-F. Shi, D. C. Ralph, J. Park,
and P. L. McEuen, Nature Phys. AOP (2012), DOI:10.1038/nphys2493.

\bibitem{Levitov2011} J. C. W. Song, M. S. Rudner, C. M. Marcus, and L. S.
Levitov, Nano Lett. \textbf{11}, 4688 (2011).

\bibitem{Jun2010} J. Yan and M. S. Fuhrer, Nano Lett. \textbf{10}, 4521 (2010).

\bibitem{McEuen2010} X. Xu, N. M. Gabor, J. S. Alden, A. M. van der Zande,
and P. L. McEuen, Nano Lett. \textbf{10}, 562 (2010).

\bibitem{Efetov2010} D. K. Efetov and P. Kim, Phys. Rev. Lett. \textbf{105}, 256805
(2010).

\bibitem{Yan2008} J. Yan, E. A. Henriksen, P. Kim, and A. Pinczuk, Phys.
Rev. Lett. \textbf{101}, 136804 (2008).

\bibitem{Martin2008} J. Martin, N. Akerman, G. Ulbricht, T. Lohmann, J. H.
Smet, K. von Klitzing, and A. Yacoby, Nature Phys. \textbf{4}, 144
(2008).

\bibitem{Levitov2012} J. C. W. Song, M. Y. Reizer, and L. S. Levitov, Phys.
Rev. Lett. \textbf{109}, 106602 (2012).

\bibitem{Zhang2008} L. M. Zhang, Z. Q. Li, D. N. Basov, M. M. Fogler, Z. Hao,
and M. C. Martin, Phys. Rev. B \textbf{78}, 235408 (2008).

\end{thebibliography}
\end{document}